# First-principles prediction of extraordinary thermoelectric efficiency in superionic Li$_2$Sn$X_3$($X$=S,Se)

Enamul Haque*[1], Claudio Cazorla**[2], and M. Anwar Hossain[1]

[1]Department of Physics, Mawlana Bhashani Science and Technology University

Santosh, Tangail-1902, Bangladesh

[2]School of Materials Science and Engineering, University of New South Wales Australia, Sydney, New South Wales 2052, Australia

*Email:enamul.phy15@yahoo.com, c.cazorla@unsw.edu.au

## Abstract

Thermoelectric materials create an electric potential when subject to a temperature gradient and *vice versa* hence they can be used to harvest waste heat into electricity and in thermal management applications. However, finding highly efficient thermoelectrics with high figures of merit, $zT \geq 1$, is very challenging because the combination of high power factor and low thermal conductivity is rare in materials. Here, we use first-principles methods to analyze the thermoelectric properties of Li$_2$Sn$X_3$ ($X$=S,Se), a recently synthesized class of lithium fast-ion conductors presenting high thermal stability. In *p*-type Li$_2$Sn$X_3$, we estimate highly flat electronic valence bands that render high Seebeck coefficients exceeding 400 μVK$^{-1}$ at 700K. In *n*-type Li$_2$Sn$X_3$, the electronic conduction bands are slightly dispersive however the accompanying weak electron-acoustic phonon scattering induces high electrical conductivity. The combination of high Seebeck coefficient and electrical conductivity gives rise to high power factors, reaching a maximum of 4 mWm$^{-1}$K$^{-2}$ in *p*-type Li$_2$SnS$_3$ and 8 mWm$^{-1}$K$^{-2}$ in *n*-type Li$_2$SnSe$_3$ at 300 K. Likewise, the thermal conductivity in Li$_2$Sn$X_3$ is low as compared to conventional thermoelectric materials, 2-5 Wm$^{-1}$K$^{-1}$ at room temperature. As a result, we estimate a maximum $zT = 1.05$ in *p*-type Li$_2$SnS$_3$ at 700 K and an extraordinary 3.07 (1.5) in *n*-type Li$_2$SnSe$_3$ at the same temperature (300 K). Our findings of huge $zT$ in Li$_2$Sn$X_3$ suggest that lithium fast-ion conductors, typically employed as electrolytes in solid-state batteries, hold exceptional promise as thermoelectric materials.

# 1. INTRODUCTION

Thermoelectric (TE) materials represent an alternative to traditional power generation sources since they can transform waste heat directly into electricity [1–5]. TE materials also can be employed in solid-state refrigeration applications based on the Peltier effect, in which heat is removed via the application of electric bias [6,7]. The efficiency of TE materials is measured by a dimensionless parameter called figure of merit defined as $zT = S^2 \sigma \kappa^{-1} T$ [8,9], where $S$ represents the Seebeck coefficient, $\sigma$ the electrical conductivity, $\kappa$ the thermal conductivity, and $T$ the operation temperature. The measured efficiency of best TE materials known to date typically are $zT < 2.6$ [10–14]; for instance, SnSe exhibits a maximum $zT$ of 2.6 at $T = 923$ K [11] and at room temperature $Bi_2Te_3$ and other related alloys present record $zT$'s of 0.8-1.2 [15–19]. However, in order to develop new and scalable energy conversion and thermal management applications there is a pressing need to find highly efficient TE materials with $zT > 3$ [20].

Highly efficient TE materials should display high power factors (PF), defined as the product of the Seebeck coefficient and electrical conductivity, and low thermal conductivities. However, is quite unusual to find those two qualities simultaneously in a same material. Some TE efficiency enhancement strategies have focused on reducing $\kappa$ while keeping PF almost unchanged [21–25]. For example, some compounds containing heavy atoms like $YbFe_4Sb_3$ [26], $CoSb_3$ [27], $Mg_2Si_{1-x}Sn_x$ [28], and $Sb_2Te_3$ [29] show good thermoelectric properties primarily due to low lattice thermal conductivity $\kappa_l$. Other $zT$ enhancement strategies have focused on optimizing PF while keeping $\kappa$ almost unchanged [26-28]. For example, a distortion in the electronic density of $p$-type PbTe caused by the introduction of Ti impurities enhances significantly its PF leading to high $zT$ values of 1.5 at 773 K [30–32]. However, $S$ and $\sigma$ normally change oppositely (e.g., $S$ increases with temperature whereas the electrical conductivity is reduced, and $S$ is low at high carrier concentrations whereas $\sigma$ is high) hence systematic improvement of TE materials through PF enhancement is difficult.

Recently, Cu- and Ag-based fast-ion conductors (FIC) have been proposed as very promising TE materials [33–36]. In superionic materials, specific ions start to diffuse through the crystalline matrix above a critical temperature rendering high ionic conductivities comparable to those in liquids. Ionic diffusion produces high structural disorder in the crystal hence enhances $zT$ by lowering $\kappa$'s. Interestingly, the intrinsic high anharmonicity in FIC seems to be enough to render low $\kappa$'s even at temperatures below the superionic transition point [37,38]. Likewise, the power factor reported for the investigated FIC are significantly large (e.g., ~1 mWm$^{-1}$K$^{-2}$ in $Cu_{2-x}Se$ at room temperature [33]). But, unfortunately, FIC are prone to suffer from thermo- and electro-migration issues that limit their application as TE materials [39,40]. Nevertheless, simple strategies based on chemical doping have been shown to solve those problems by simultaneously suppressing cation migration and maintaining low lattice thermal conductivity [41,42]. Therefore, in view of the large $zT$'s measured in Cu- and Ag-based FIC and the straightforward solutions proposed for addressing their likely stability issues, it is of great interest to continue exploring the TE potential of superionic materials.

Here, we investigate the TE performance of $Li_2SnX_3$ ($X$=S, Se), a recently synthesized family of lithium superionic compounds that present high thermal stability [54-56], by using first-principles computational methods. Our calculations predict exceptional $zT$'s of 1-1.5 at room temperature and of >3 at 700 K, due to the combined effect of high PF and low κ. To the best of our knowledge, this is the first prediction of extraordinary $zT$ in a Li-based FIC that have been synthesized previously in the laboratory. Therefore, lithium superionic conductors, which are well-known for their use in electrochemical devices and currently are being investigated very actively, appear to be extremely promising TE materials.

## 2. COMPUTATIONAL METHODS

Structural relaxations and electron-phonon (e-ph) dynamical matrix calculations are performed with the plane-wave pseudopotential method as implemented in the Quantum Espresso code [43,44]. The adopted exchange-correlation scheme corresponds to the generalized gradient approximation due to Perdew-Burke-Ernzerhof (PBE) [45,46]. In these calculations Vanderbilt ultrasoft pseudopotentials [47] are employed, along with a 42 Ry cutoff energy in the wavefunctions and Methfessel-Paxton smearing [48] of 0.03 Ry. For charge density calculations, non-shifted **k**-point grids of $4 \times 2 \times 2$ and $4 \times 4 \times 4$ are used for $Li_2SnS_3$ and $Li_2SnSe_3$, respectively. For electron-phonon calculations, $1 \times 1 \times 1$ Γ-centered **q**-point grids are used since the unit cells of $Li_2SnS_3$ and $Li_2SnSe_3$ are quite large (48 and 24 atoms, respectively) and in that case coarse **q**-point grids are sufficient to generate e-ph matrices with reasonable accuracy [49]. Electron-phonon averaged approximation (EPA) calculations are performed with the EPA code [49] by using an energy grid spacing of 0.6 eV. Consistency tests were performed by repeating some calculations with the EPA-MLS code (which computes electron-phonon matrices via moving least squares and produces accurate results with coarse q-point grids) and obtaining good agreement between the two series of results (data not presented here) [50].

Electronic structure calculations are performed with the full-potential linearized augmented plane wave method as implemented in WIEN2k code [51] using the PBE and TB-mBJ (Tran-Blaha modified Becke-Johnson) exchange-correlation functionals [52]. 2,600 irreducible **q**-points and a $R_{mt}K_{max}$ product of 8.0 (where $R_{mt}$ is the muffin-tin sphere radius and $K_{max}$ the plane wave cut-off energy) are used to calculate electronic band structure and density of states. $28 \times 28 \times 12$ and $22 \times 22 \times 17$ non-shifted **q**-point grids are used for the self-consistent calculation of energy eigenvalues, which are required for transport calculations, in $Li_2SnS_3$ and $Li_2SnSe_3$, respectively. The calculation of transport coefficients is performed with the modified BoltzTrap code [49,53], in which the carrier relaxation time is calculated trough the equation [49]

$$\tau^{-1}(\epsilon, \mu, T) = \frac{2\pi\Omega}{g_s\hbar} \sum_v \{g_v^2(\epsilon, \epsilon + \bar{\omega}_v)[n(\bar{\omega}_v, T) + f(\epsilon + \bar{\omega}_v, \mu, T)] \rho(\epsilon + \bar{\omega}_v) \\ + g_v^2(\epsilon, \epsilon - \bar{\omega}_v)[n(\bar{\omega}_v, T) + 1 - f(\epsilon - \bar{\omega}_v, \mu, T)]\rho(\epsilon - \bar{\omega}_v)\} \dots \dots (1)$$

where Ω is the volume of the primitive unit cell, ℏ the reduced Planck's constant, $v$ the phonon mode index, $\bar{\omega}_v$ the averaged phonon mode energy, $g_v^2$ the averaged electron-phonon matrix, $n(\bar{\omega}_v, T)$ the Bose-Einstein distribution function, $f(\epsilon + \bar{\omega}_v, \mu, T)$ the Fermi-Dirac distribution function, $g_s = 2$ the spin degeneracy, $\epsilon$ the electron energy, and ρ the density of states per unit energy and unit volume. Further details on the calculation of transport coefficients can be found in work [49].

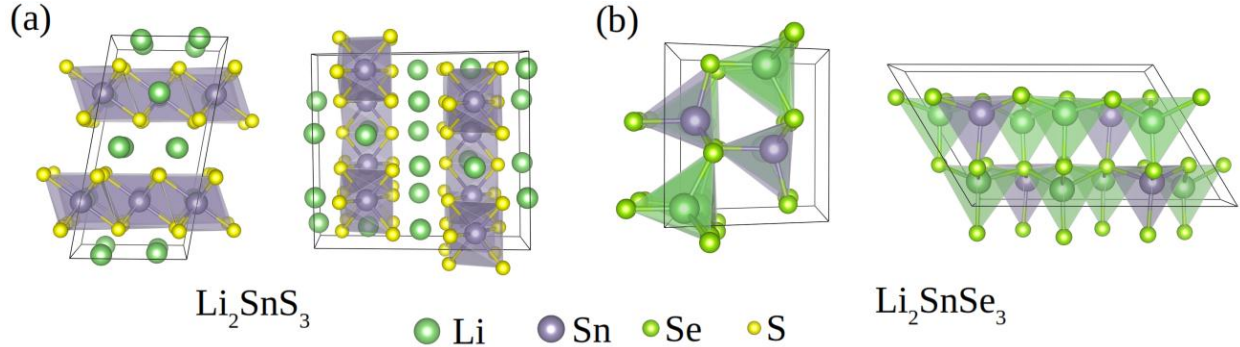

Fig. 1. Ground state crystal structure of (a) Li$_2$SnS$_3$ (monoclinic symmetry, space group *C2/c*) and (b) Li$_2$SnSe$_3$ (monoclinic symmetry, space group *Cc*) considering different views. The black lines represent the corresponding unit cells.

## 3. RESULTS AND DISCUSSION

Figure 1 shows the crystal structure of Li$_2$SnX$_3$ (X=S, Se). Li$_2$SnS$_3$ crystallizes in a monoclinic phase with space group *C2/c* (#15) [54,55] and Li$_2$SnSe$_3$ in another monoclinic phase with space group *Cc* (#9) [56]. The number of atoms per unit cell is 48 and 24 for Li$_2$SnS$_3$ and Li$_2$SnSe$_3$, respectively. The corresponding fully optimized lattice parameters are listed in Table I, which are in good agreement with the available experimental data and previous theoretical work done by other authors.

Table I. Fully relaxed lattice parameters in Li$_2$SnX$_3$ (X=S, Se). Available experimental data are shown for comparison. SC-XRD stands for single crystal x-ray diffraction and SXRPD for synchrotron X-ray powder diffraction. Further structural data can be found in the Supplementary Material.

| Compounds | a | b | c | β | Ref. |
|---|---|---|---|---|---|
| Li$_2$SnS$_3$ | 6.4542 | 11.1853 | 12.4891 | 99.884 | This |
| | 6.3964 | 11.0864 | 12.405 | 99.867 | Expt.(SC-XRD) [54] |
| | 6.3961 | 11.089 | 12.416 | 99.860 | Expt.(SC-XRD) [55] |
| | 6.4004 | 11.0854 | 12.4222 | 99.883 | Expt.(SXRPD) [54] |
| | 6.30 | 10.91 | 12.15 | 99.94 | LDA [57] |
| | 6.28 | 10.87 | 12.08 | 99.82 | LDA [54] |
| Li$_2$SnSe$_3$ | 12.7363 | 7.3198 | 7.9506 | 121.14 | This |
| | 12.522 | 7.2137 | 7.7692 | 120.96 | Expt.(SC-XRD) [56] |

We note that PBE-GGA overestimates the experimental lattice parameters by less than 1% while LDA underestimates them by ~2%. In view of the superior agreement with experiments provided by PBE-GGA, we will use this functional in the rest of calculations if not stated otherwise. Further structural data obtained in $Li_2SnX_3$ (fractional atomic coordinates) can be found in Supplementary Tables S1.

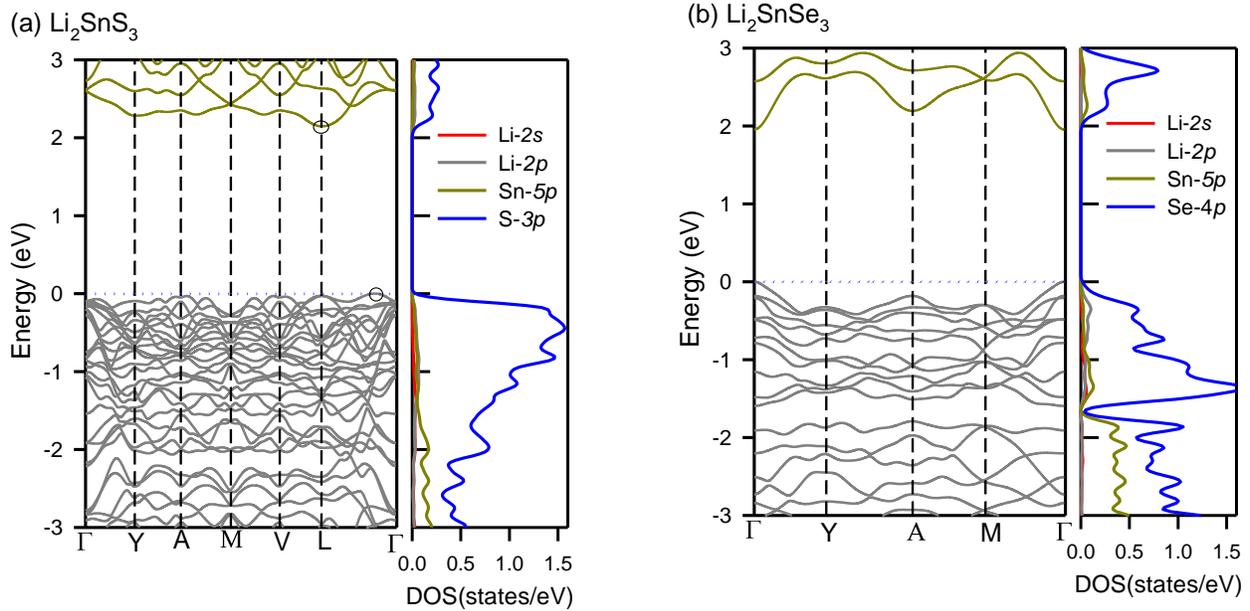

Fig. 2. Electronic band structure and projected density of states of (a) $Li_2SnS_3$ and (b) $Li_2SnSe_3$. The dashed blue line at zero energy represents the Fermi level. The K-point path of the calculation for the monoclinic $Li_2SnS_3$ and $Li_2SnSe_3$ (generated by Xcrysden program [58,59]) starts at $\Gamma(0,0,0)$ and goes through $Y(1/2,1/2,0)$, $A(0,0,1/2)$, $M(1/2,1/2,1/2)$, $V(1/2,0,0)$, and $L(1/2,0,1/2)$ high symmetry points, ending at $\Gamma$.

The electronic band structure and electronic density of states in $Li_2SnX_3$ as computed with a combination of the PBE-GGA and TB-mBJ functionals, are shown in Fig. 2. Both compounds display similar electronic band structure properties. In $Li_2SnS_3$, the maxima of the valence band (VBM) lies between the L and $\Gamma$ points while the minima of the conduction band (CBM) sits at L. An indirect band gap of 2.14 eV is estimated in $Li_2SnS_3$, in consistent agreement with previous theoretical calculations and the experimental value of 2.38 eV [54]. A direct band gap with almost the same value exists at the L point, hence $Li_2SnS_3$ may behave also as a direct band gap semiconductor. In $Li_2SnSe_3$, both CBM and VBM lie at the $\Gamma$-point and render a direct band gap of 1.95 eV. The energy bands in both compounds are highly flat which, as we will explain in detail below, leads to high Seebeck coefficients (according to Mott's relation [30]) and high power factors. The total and projected density of states in both compounds are shown in the left panels of each band structure figure. Orbitals Sn $5p$ and S $3p$-Se $4p$ are strongly hybridized and contribute the most to VBM and CBM in Li$_2$SnX$_3$ (X=S, Se), which suggests that doping in the Sn-site or S/Se site, or co-doping, could be used to tune further the band gap.

Phonons characterize the dynamical stability of a compound; real frequencies in the phonon dispersion curves indicate that the crystal is vibrationally stable. Elastic stability, on the other hand, can be assessed from the material elastic constants. The phonon band structure calculated in Li$_2$SnX$_3$ is shown in Fig.3. As can be appreciated in there, all lattice phonon modes and frequencies are well-behaved (i.e., are real). The elastic constants computed in both compounds are listed in Table II, along with the corresponding Pugh's ratio [60], Poisson's ratio, and Debye temperature.

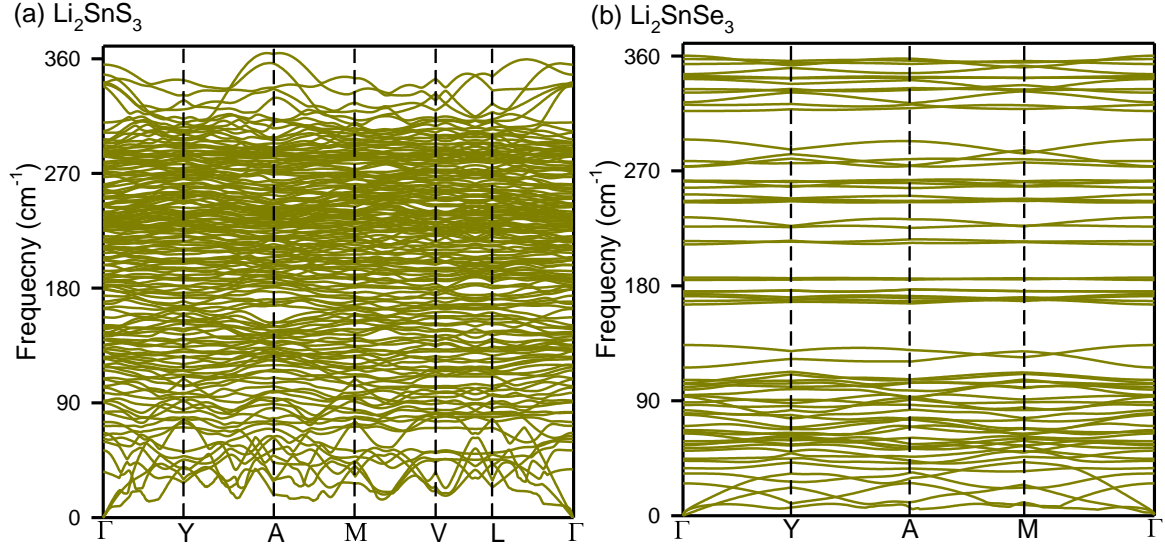

Fig. 3. Phonon dispersion relations of (a) Li$_2$SnS$_3$ and (b) Li$_2$SnSe$_3$.

Table II. Calculated elastic constants ($c_{ij}$), Pugh's ratio (B/G), Poisson's ratio (ν), longitudinal and transverse sound velocity ($v_l$ & $v_t$), and Debye temperature ($\theta_D$).

| Parameter | Li$_2$SnS$_3$ | Li$_2$SnSe$_3$ |
|---|---|---|
| $c_{11}$ | 90.24 | 34.89 |
| $c_{12}$ | 29.36 | 15.08 |
| $c_{13}$ | 31.06 | 13.10 |
| $c_{16}$ | -4.29 | 0.14 |
| $c_{22}$ | 86.64 | 46.39 |
| $c_{23}$ | 16.40 | 13.02 |
| $c_{26}$ | 3.78 | 1.45 |
| $c_{33}$ | 89.28 | 38.42 |
| $c_{36}$ | 8.16 | -1.92 |
| $c_{44}$ | 24.75 | 11.81 |
| $c_{45}$ | 2.39 | -0.10 |
| $c_{55}$ | 33.75 | 13.57 |
| $c_{66}$ | 22.28 | 13.48 |
| B/G | 1.66 | 1.73 |
| ν | 0.249 | 0.257 |
| $v_l$ | 4933.76 | 3195.14 |
| $v_t$ | 2850.49 | 1825.55 |
| $v_m$ | 3164.39 | 2028.54 |
| $\theta_D$ | 356.184 | 202.742 |

The elastic constants in Li$_2$SnX$_3$ fulfill the elastic stability criteria corresponding to monoclinic crystals as reported in Ref [61]. The value of the Pugh's ratio is very close to the critical value of 1.75 that differentiates ductile from brittle materials. The low value of the Debye temperature indicates low heat transport by phonons, that is, low lattice thermal conductivity, $\kappa_l$. In order to calculate $\kappa_l$, we use the longitudinal ($v_l$) and transverse ($v_t$) sound velocities to estimate the acoustic Gruneisen parameter [62]:

$$\gamma_a = \frac{9(v_l^2 - 4v_t^2/3)}{2(v_l^2 + 2v_t^2)} \dots \dots \dots \dots (2)$$

and the Poisson's ratio, ν, to estimate the elastic Gruneisen parameter [63]:

$$\gamma_e = \frac{3}{2}\left(\frac{1+\nu}{2-3\nu}\right) \dots \dots \dots \dots (3)$$

The total Gruneisen parameter, γ, which is the sum of $\gamma_a$ and $\gamma_e$, then can be used in the Slack equation to calculate the lattice thermal conductivity as [64,65]:

$$\kappa_l = A \frac{M_{av}\theta_{aco}^3 \delta}{\gamma^2 n^{2/3} T} \dots \dots \dots \dots \dots (4)$$

where $M_{av}$ is the average atomic mass, δ the cubic root of the volume of the primitive cell, $\theta_{aco}$ the acoustic Debye temperature, $n$ the number of atoms per unit cell, and $A$ the coefficient defined by the relation:

$$A(\gamma) = \frac{5.720 \times 10^7 \times 0.849}{2 \times \left[1 - \left(\frac{0.514}{\gamma}\right) + \left(\frac{0.228}{\gamma^2}\right)\right]} \dots \dots \dots (5)$$

Equation 4 provides an averaged isotropic value for $\kappa_l$, however in most materials lattice thermal conductivity depends on the direction of heat propagation. Equation 4 can be modified as follows to reproduce likely anisotropic effects in the lattice thermal conductivity:

$$\kappa_l = A \frac{M_{av}\theta_i^3 \delta}{\gamma^2 n^{2/3} T} \quad (i = a, b, c) \dots \dots \dots (6)$$

where $\theta_i$ is the anisotropic Debye temperature, which can be calculated with the equation [66]:

$$\theta_i = \frac{h}{k_B}\left(\frac{3n}{4\pi}\frac{N_A\rho}{M}\right) v_{mi}, \quad i = a, b, c. \ldots\ldots\ldots (7)$$

In Eq. (7), $v_{mi}$ represents the average speed of sound along the crystallographic direction a, b, and c. By using the elastic constants and density of the material, $v_{mi}$ can be calculated via the Christoffel eigenvalue equation [67–69]:

$$\sum_{ijkl}|\rho v^2(\boldsymbol{n})\delta_{ij} - c_{ijkl}\boldsymbol{n}_k\boldsymbol{n}_l| = 0 \ldots\ldots\ldots (8)$$

where $\rho$ is the density of the compound, $v^2(\boldsymbol{n})$ the speed of sound along with the unit vector $\boldsymbol{n}$, $c_{ijkl}$ the fourth rank elastic tensor, and $\boldsymbol{n}_k$, $\boldsymbol{n}_l$ are unit vector components representing the direction of propagation (one longitudinal and two transverse) [70].

Table III: Calculated anisotropic sound velocity ($v_{mi}$), Debye temperature ($\theta_i$) and averaged Gruneisen parameter ($\gamma_{av}$) of $Li_2SnX_3$

| Compound | $v_{ma}$ (m/s) | $v_{mb}$ (m/s) | $v_{mc}$ (m/s) | $\theta_a$ (K) | $\theta_b$ (K) | $\theta_c$ (K) | $\gamma_{av}$ |
|---|---|---|---|---|---|---|---|
| $Li_2SnS_3$ | 2607.4 | 2839.8 | 2345.8 | 254.1 | 276.8 | 228.6 | 1.67 |
| $Li_2SnSe_3$ | 1778.3 | 1754.3 | 1617.8 | 177.7 | 175.3 | 161.7 | 1.72 |

The anisotropic Debye temperatures and wave velocities estimated along the corresponding crystallographic directions in $Li_2SnX_3$ are shown in Table III. The average of the anisotropic Debye temperatures is in good agreement with the isotropic Debye temperature (see Table II), which supports the validity of the computational $\kappa_l$ method employed in this study. Meanwhile, the calculated large Gruneisen parameters indicate high anharmonicity in the crystals and hence intense phonon scattering. By using Eq. (6), we estimate the lattice thermal conductivity along the three crystallographic directions and represent them in Fig.4. Since $Li_2SnSe_3$ undergoes a structural phase transition around 750 K [40], is reasonable to calculate $\kappa_l$ just within the temperature interval 200-700K. In both compounds, non-negligible anisotropies in thermal transport are observed and $\kappa_l$ is lowest along the c-axis.

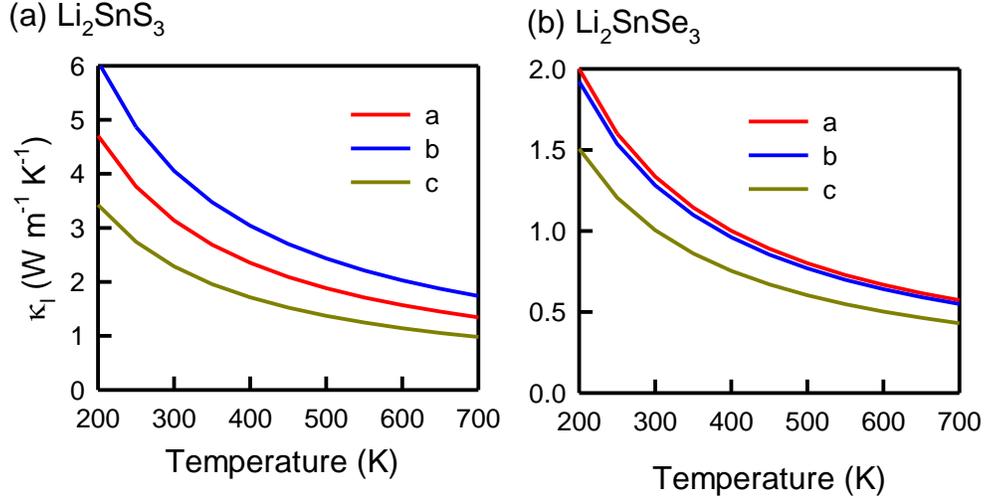

Fig. 4. Anisotropic lattice thermal conductivity of (a) Li$_2$SnS3 and (b) Li$_2$SnSe$_3$. The symbols a, b, and c, indicates three crystallographic directions (x, y, z).

Li$_2$SnS$_3$ presents larger $\kappa_l$ than Li$_2$SnSe$_3$, essentially due to higher anisotropic Debye temperatures (Table III). For example, at 300K the lattice thermal conductivity along the b-axis is equal to 4 W/m K in Li$_2$SnS$_3$ and to 1.5 W/m K in Li$_2$SnSe$_3$. We note also that the $\kappa_l$ maximum in Li$_2$SnS$_3$ is obtained along the b-axis whereas in Li$_2$SnSe$_3$ along the a-axis. This lattice thermal conductivity behavior is similar to the thermal transport trends observed in SnS and SnSe [71]. For example, the average $\kappa_l$ calculated in Li$_2$SnSe$_3$ is very close to that in SnSe (~1 W/m K at 300 K) [71], although the lattice thermal conductivity in Li$_2$SnS$_3$ is much higher than that in SnS (due the considerable mass difference between Li and Se as compared to Li and S). It is worth noting that the $\kappa_l$ estimated in Li$_2$Sn$X_3$ are low as compared to those reported in other typical thermoelectric materials like CoSb$_3$ (11.5 W/m K at 300 K) [27].

In Boltzmann transport theory, the constant relaxation approximation can be used to calculate electrical conductivities and the electronic contribution to the thermal conductivity. Previous theoretical calculations, however, have shown that such approximation may lead to some inaccuracies [47] since the carrier relaxation time, $\tau$, depends strongly on the carrier density and composition of the material. Here, we employ a more sophisticated method to estimate carrier relaxation times that is able to take into account electron-phonon interactions with reasonable accuracy (Eq.1 in Computational Methods section) [49]. Figure 5 encloses the $\tau$ results obtained at 300K in Li$_2$Sn$X_3$ as a function of energy.

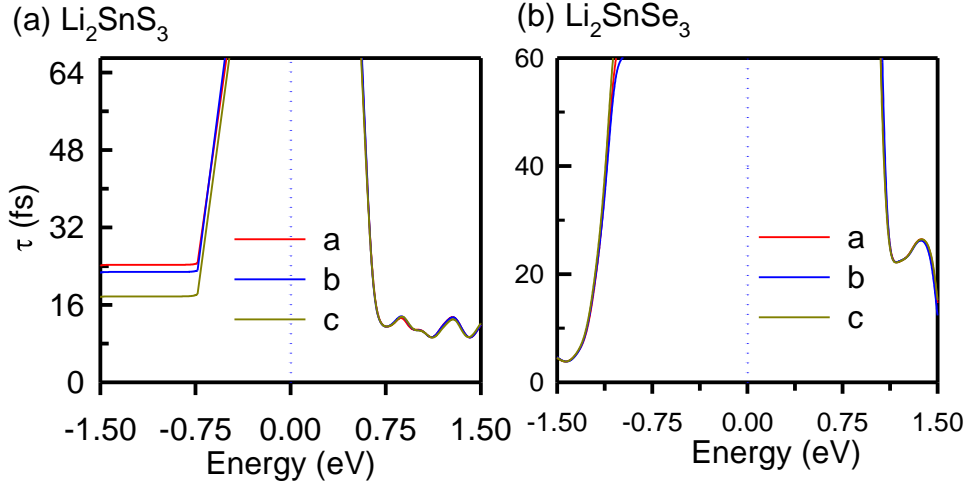

Fig. 5. Computed carrier lifetime (at 300 K) of (a) Li$_2$SnS3 and (b) Li$_2$SnSe$_3$. The dotted blue line at zero energy represents the Fermi level.

Both compounds present practically isotropic relaxation times. In Li$_2$SnS$_3$, the relaxation time of electrons in conduction bands is almost the same than that of holes in valence bands. By contrast, in Li$_2$SnSe$_3$ the electrons in conduction bands near the Fermi energy level have longer relaxation times than the holes in valence bands. The slightly dispersive nature of the conduction bands, as compared to the valence bands near the Fermi level, is responsible for the τ behavior observed in Li$_2$SnSe$_3$. In both compounds, τ decreases sharply near the band edges. This effect can be rationalized with the help of the relationship [49]:

$$\tau^{-1} \sim g^2(\epsilon)\rho(\epsilon) \dots \dots \dots (9)$$

which points out that the relaxation time varies inversely with the carrier density of states per unit energy and volume, $\rho$, when the electron-phonon matrix elements, $g$, depend just weakly on the energy.

Above the Debye temperature, electron-phonon (e-ph) scattering is dominant [72,73] and the e-ph coupling matrix alone can be used to describe with reasonable accuracy τ, the electrical conductivity, and the electronic part of the thermal conductivity. Here, we are interested in the temperature interval 300-700K, which is above $\theta_i$ (Table III), hence we take into consideration just the e-ph coupling matrix for the calculation of electronic transport coefficients. Figure 6 encloses the Seebeck coefficients calculated in Li$_2$SnX$_3$ as a function of carrier concentration at temperatures $T = 300$, 500, and 700K. Both compounds show practically isotropic Seebeck coefficients, which are larger than 400 µV/K at 700 K and carrier concentrations of ~$10^{19}\ cm^{-3}$ in p-type systems. The Seebeck coefficient corresponding to p-type Li$_2$SnX$_3$ is much larger than that corresponding to *n*-type due to the relatively flatter valence bands calculated in the former cases. We also note that the Seebeck coefficient calculated in Li$_2$SnS$_3$ is much larger than in Li$_2$SnSe$_3$ due to the smaller band gap estimated in the latter case. The Seebeck coefficients in *p*-type Li$_2$SnS$_3$ and Li$_2$SnSe$_3$ are very similar to those in *p*-type SnS and SnSe, respectively [71], although the Seebeck coefficients in *n*-type Li$_2$SnX$_3$ are slightly smaller.

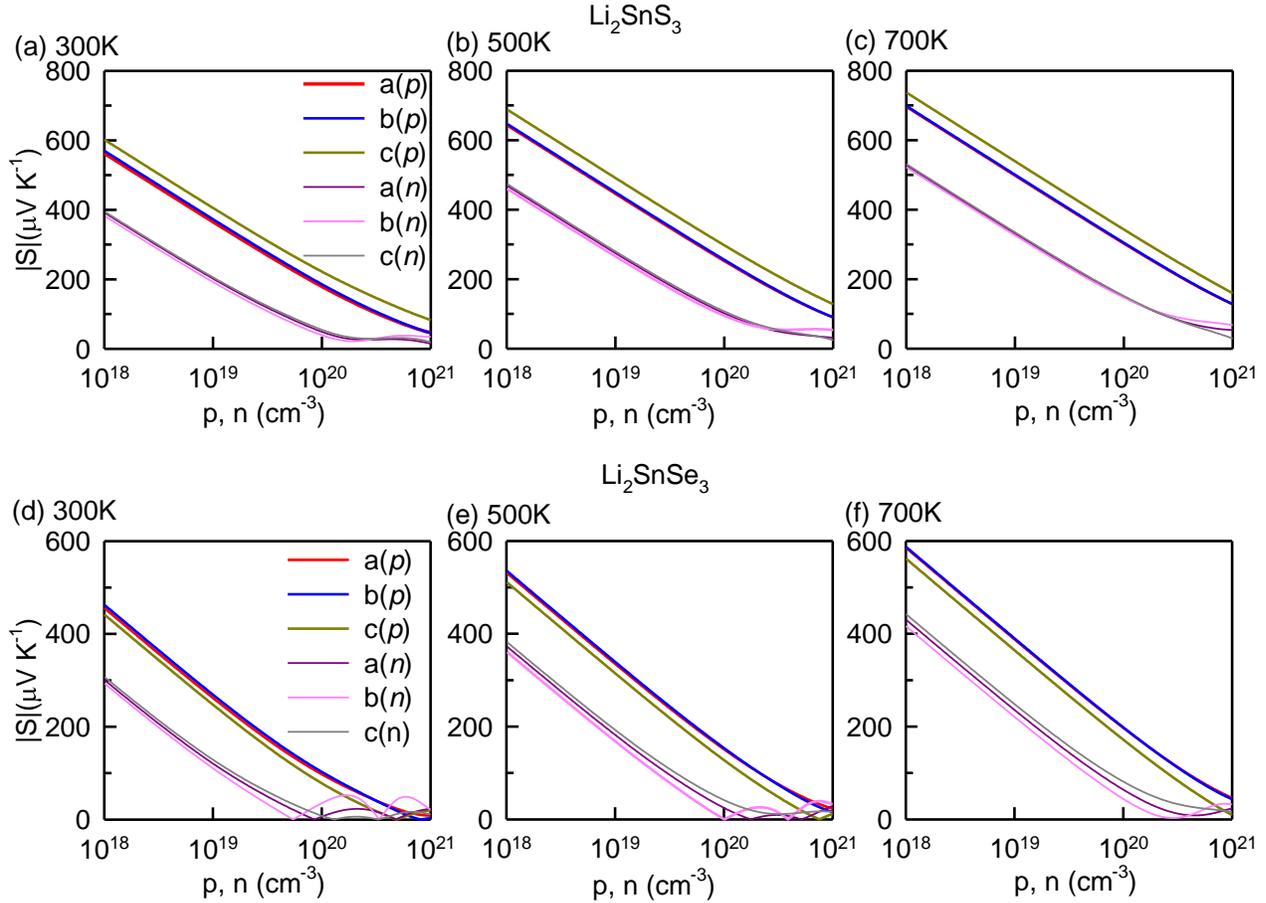

Fig. 6. Calculated absolute value of Seebeck coefficient as a function of carrier concentration of Li$_2$SnS$_3$ (top panel) and (b) Li$_2$SnSe$_3$ (bottom panel). The symbols a, b, and c denote the crystallographic direction while p and n indicate the type of the carrier.

The high Seebeck coefficients calculated in Li$_2$SnX$_3$ lead to high power factors (PF), as is shown in Fig.7. The estimated PF display non-isotropic behavior due to the anisotropies found in the electrical conductivity (Supplementary Material). The energy band gap in both Li-based compounds are almost two times larger than those found in SnS and SnSe, however their electrical conductivities are much higher as well. In particular, Li ions render low effective masses ($m_e^* = 0.13m_0, m_h^* = 0.14m_0$ and $m_e^* = 0.04m_0, m_h^* = 0.09m_0$ in Li$_2$SnS$_3$ and Li$_2$SnSe$_3$, respectively) as compared to SnS ($m_e^* = 0.20m_0, m_h^* = 0.28m_0$ [74]) and SnSe ($m_e^* = 0.12m_0, m_h^* =$

$0.13m_0$ [75]) thus inducing higher electrical conductivity in $Li_2SnX_3$. The electrical conductivity in p-type $Li_2SnS_3$ is much higher than in the n-type case due to the lower effective mass of holes. By contrast, the electrical conductivity in n-type $Li_2SnSe_3$ is much higher than in the p-type case due to the lower effective mass of electrons. Actually, the PF in *p*-type $Li_2SnS_3$ along the a- and b-axes can reach large values of ~4 mW m$^{-1}$K$^{-2}$ at room temperature (Fig.7). Moreover, the PF estimated in $Li_2SnSe_3$ at 300 K amounts to ~8 mW m$^{-1}$K$^{-2}$ which is one of the highest PF predicted to date in *n*-type semiconducting materials (PF measurements of 8 mW m$^{-1}$K$^{-2}$ in *p*-type half-Heusler NbFeSb at 500 K set the current experimental record [69]).

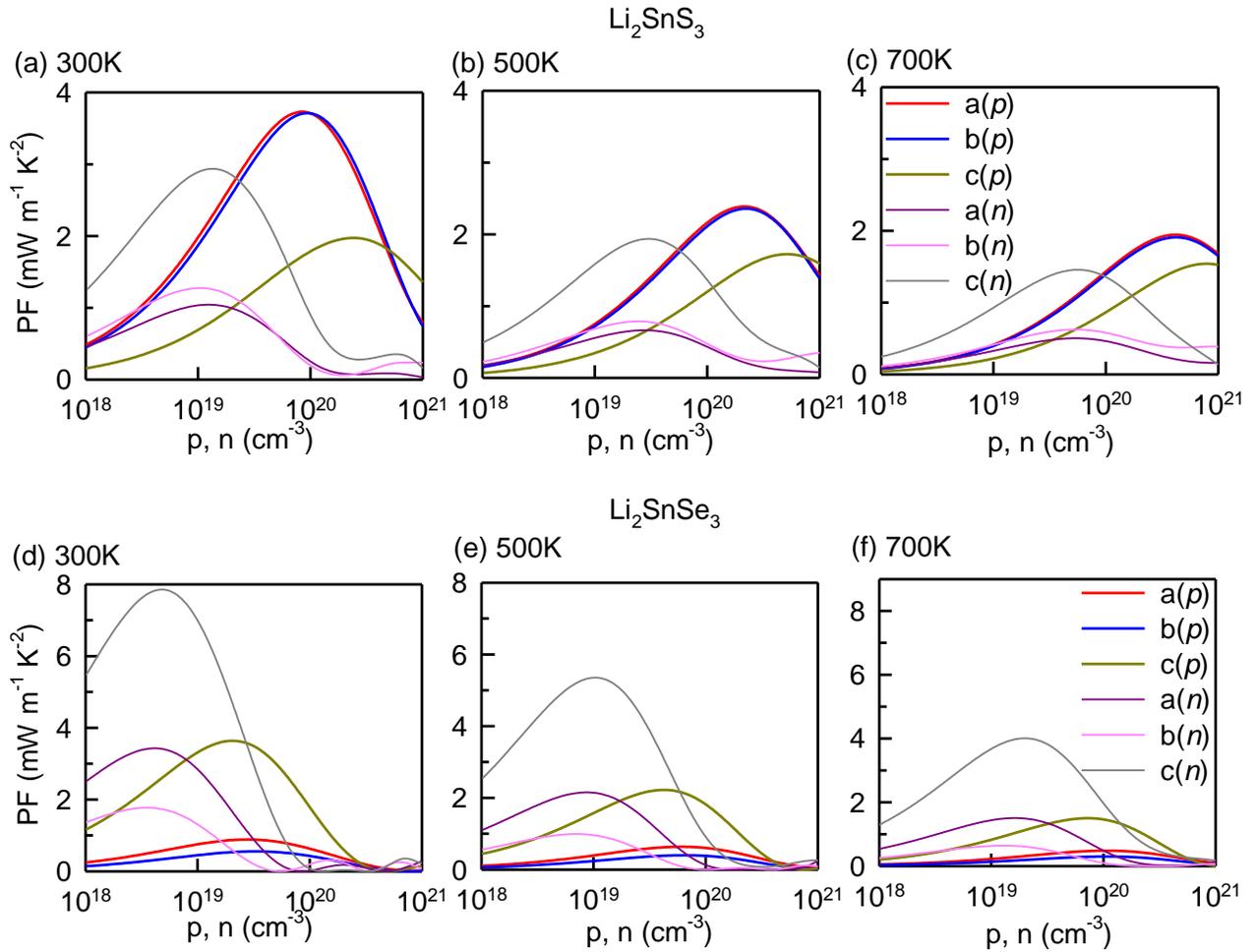

Fig. 7. Computed power factor ($PF = S^2\sigma$) as a function of carrier concentration of $Li_2SnS3$ (top panel) and (b) $Li_2SnSe_3$ (bottom panel).

Figure 8 encloses the total thermal conductivity, equal to the sum of the lattice and electronic contributions, calculated in $Li_2SnX_3$ at different temperatures and expressed as a function of carrier concentration. Like SnS and SnSe, the thermal conductivity in both compounds show anisotropic behavior. At low temperature and low carrier concentration the lattice contribution to the thermal conductivity is dominant in both compounds, whereas at high carrier concentration (above $\sim 10^{20}\ cm^{-3}$) the electronic contribution to the thermal conductivity is dominant (further details on the electronic part of the thermal conductivity can be found in Supplementary Fig.2). The $\kappa$ estimated in these compounds are lower than in other typical semiconductors (e.g., 11.5 W/m K in CoSb$_3$ at 300 K [27]), although higher than in SnS and SnSe due to larger lattice thermal conductivities. Such estimated low thermal conductivities can potentially lead to high thermoelectric figures of merit.

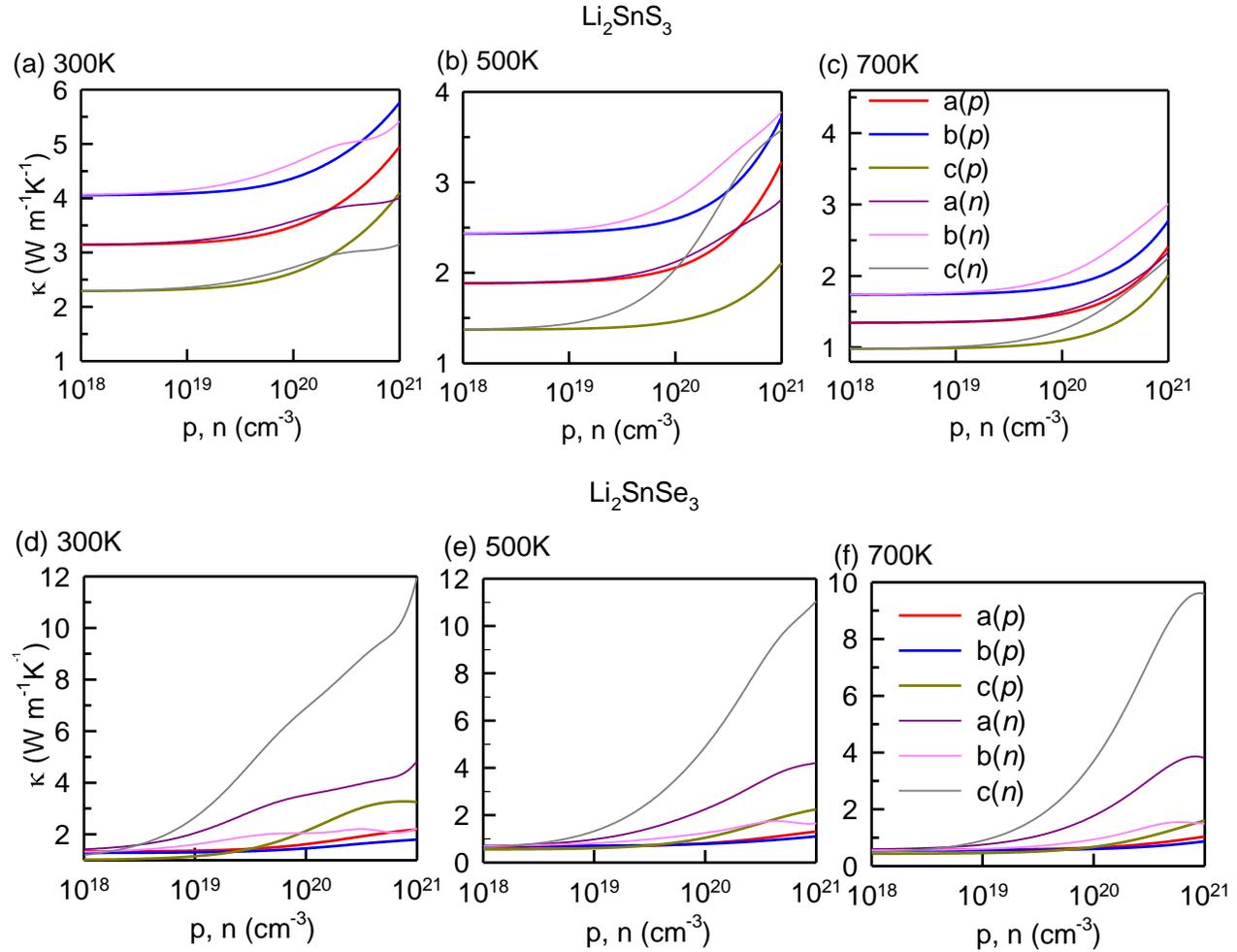

Fig. 8. Total thermal conductivity ($\kappa = \kappa_e + \kappa_l$) of Li$_2$SnS$_3$ (top panel) and (b) Li$_2$SnSe$_3$ (bottom panel) as a function of carrier concentration.

Figure 9 shows the $zT$ results obtained in Li$_2$SnX$_3$ expressed as a function of carrier concentration and temperature. At room temperature, the $zT$ estimated in Li$_2$SnS$_3$ is relatively low, 0.1-0.4, due to its relatively high thermal conductivity obtained at low temperatures (Fig.8). Nevertheless, $zT$ increases significantly under increasing temperature; for instance, at 500 K the $zT$ in $n$-type Li$_2$SnS$_3$ along the c-axis amounts to 0.65 at a carrier concentration of $1.77 \times 10^{19}\ cm^{-3}$. Yet, this value is lower than found, for instance, in $n$-type SnS ($zT$=1.5 at 750K and carrier concentration of $\sim 8 \times 10^{19}\ cm^{-3}$ [71]). The maximum $zT$ estimated in p-type Li$_2$SnS$_3$ is 1.05, which corresponds to the c-axis at 700 K and carrier concentration of $1.22 \times 10^{20}\ cm^{-3}$.

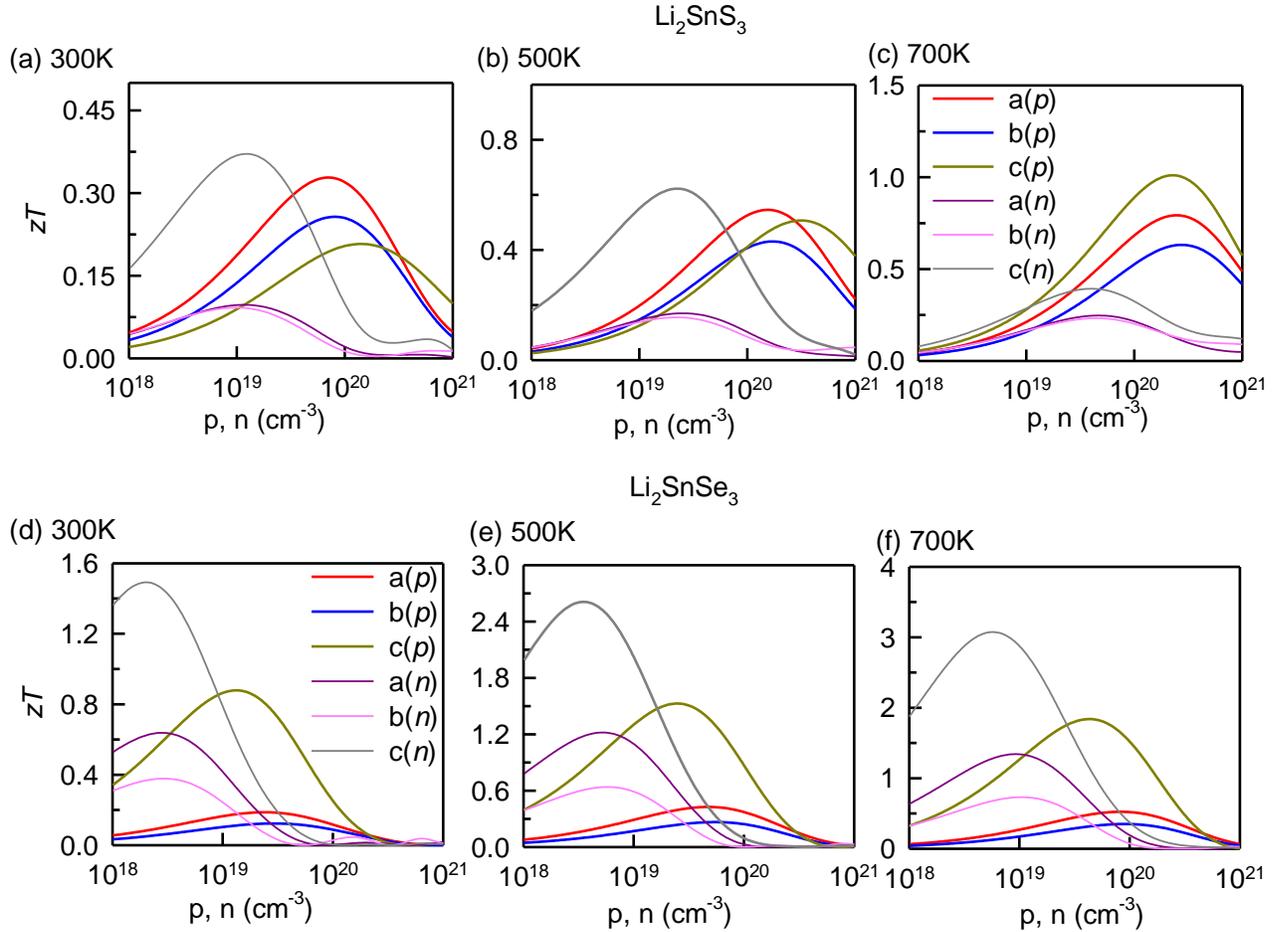

Fig. 9. Dimensionless thermoelectric figure of merit ($zT$) of Li$_2$SnS$_3$ (top panel) and (b) Li$_2$SnSe$_3$ (bottom panel) as a function of carrier concentration.

By contrast, the combination of low the thermal conductivity and high power factor leads to impressive $zT$'s in $n$-type Li$_2$SnSe$_3$. At 300K, our calculations predict a maximum $zT$ of 1.5 along the c-axis and carrier concentration $2 \times 10^{18}\ cm^{-3}$. Therefore, Li$_2$SnSe$_3$ represents a promising room-temperature thermoelectric material. Remarkably, the maximum $zT$ calculated in $n$-type Li$_2$SnSe$_3$ amounts to 3.07 at 700 K (along the c-axis and for a carrier concentration of $5.99 \times 10^{18}\ cm^{-3}$), which represents an unusually large TE figure of merit. In fact, such a $zT$ value to the best of our knowledge is the highest that has been predicted to date in an experimentally synthesized bulk material. (We note that recently a giant $zT$ of 5 has been predicted in the half-Heusler Ba$_2$BiAu at 800K [76], however that compound has not been synthesized yet in the laboratory.) It is worth noting that Li, Sn, and Se are all abundant and cost-effective elements, hence thermoelectric compounds containing them are particularly attractive from an applied point of view.

## 4. CONCLUSIONS

In summary, the thermoelectric properties of superionic $Li_2SnX_3$ ($X$=S,Se) have been studied extensively by using accurate first-principles methods. Highly flat valence bands in *p*-type $Li_2SnX_3$ lead to high Seebeck coefficients, which can exceed 400 μV/K at 700K. Slightly dispersive conduction bands induce comparatively lower Seebeck coefficients in *n*-type compounds and longer electron lifetime near the band edges. Such conduction bands lead to weak electron-phonon interactions that render high electrical conductivity in *n*-type $Li_2SnX_3$. The combination of high Seebeck coefficient and electrical conductivity gives rise to high power factors, which at room temperature can reach 4 mW m$^{-1}$K$^{-2}$ in *p*-type $Li_2SnS_3$ and 8 mW m$^{-1}$K$^{-2}$ in *n*-type $Li_2SnSe_3$. Likewise, the thermal conductivities estimated in $Li_2SnX_3$ are low, for instance, 2-5 Wm$^{-1}$K$^{-1}$ at 300 K. Based on these parameters, we predict that *p*-type $Li_2SnS_3$ is a potentially good TE material at high temperatures ($zT$=1.05 at 700 K) whereas *n*-type $Li_2SnSe_3$ is a potentially superb TE material even at room temperature ($zT$=1.5 at 300 K and 3.07 at 700K). The reported huge figures of merit suggest that lithium-based fast-ion conductors hold tremendous promise for energy conversion and thermal management applications, thus the field of thermoelectricity could benefit immensely from the intensive research already undertaken on electrochemical energy devices. Hence, we hope that the theoretical results presented in this study will motive experimentalists to investigate the TE performance of $Li_2SnX_3$ ($X$=S,Se) and other similar Li-based superionic materials.

The author thanks to Professor Boris Kozinsky, and Daehyun Wee for sharing their personal codes and Dr. K.C. Bhamu for helpful discussion about EPA calculation.

# Supplementary Material

# First-principles prediction of extraordinary thermoelectric efficiency in superionic $Li_2SnX_3$ ($X$=S,Se)

Enamul Haque*[1], Claudio Cazorla**[2], and M. Anwar Hossain[1]

[1]Department of Physics, Mawlana Bhashani Science and Technology University

Santosh, Tangail-1902, Bangladesh

[2]School of Materials Science and Engineering, University of New South Wales Australia, Sydney, New South Wales 2052, Australia

*Email: enamul.phy15@yahoo.com, c.cazorla@unsw.edu.au

## S. 1. Structural properties

The fully optimized fractional atomic coordinates of both compounds are listed in Table S1. The computed values are in well agreement with experimental values.

Table S1: Optimized fractional atomic coordinates of $Li_2SnX_3$

| Compound | Atom | Site | Calc. (PBE) | Experimental [1,2] |
|---|---|---|---|---|
| $Li_2SnS_3$ | Li | 8f | (0.2411, 0.0860, 0.9998) | (0.2466, 0.0840, 0.9997) |
|  | Li | 4e | (0.0000, 0.4168, 0.2500) | (0.0000, 0.4173, 0.2500) |
|  | Li | 4d | (0.2500, 0.2500, 0.5000) | (0.2500, 0.2500, 0.5000) |
|  | Sn | 4e | (0.0000, 0.0832, 0.2500) | (0.0000, 0.0833, 0.2500) |
|  | Sn | 4e | (0.0000, 0.7495, 0.2500) | (0.0000, 0.7499, 0.2500) |
|  | S | 8f | (0.1126, 0.0830, 0.6298) | (0.1115, 0.0831, 0.6312) |
|  | S | 8f | (0.1345, 0.2426, 0.1296) | (0.1355, 0.2418, 0.1311) |
|  | S | 8f | (0.3657, 0.0917, 0.3723) | (0.3651, 0.0920, 0.3708) |
| $Li_2SnSe_3$ | Li | 4a | (0.5573, 0.9372, 0.1502) | (0.5610, 0.9350, 0.1600) |
|  | Li | 4a | (0.3823, 0.4102, 0.1202) | (0.3720, 0.4110, 0.1030) |
|  | Sn | 4a | (0.2408, 0.9152, 0.1554) | (0.2428, 0.9165, 0.1571) |
|  | Se | 4a | (0.1008, 0.0787, 0.2660) | (0.1031, 0.0776, 0.2685) |
|  | Se | 4a | (0.4341, 0.0974, 0.2763) | (0.4349, 0.0993, 0.2788) |
|  | Se | 4a | (0.2653, 0.5834, 0.2499) | (0.2668, 0.5857, 0.2508) |

## S. 2. Transport properties

Fig. S1. Demonstrates carrier concentration dependent anisotropic electrical conductivity of $Li_2SnS_3$ (top panel) and $Li_2SnSe_3$ (bottom panel). Electrical conductivity of both compounds shows highly anisotropic behavior. For n-type carrier in both compounds the electrical conductivity along c-axis is higher compared to other axes.

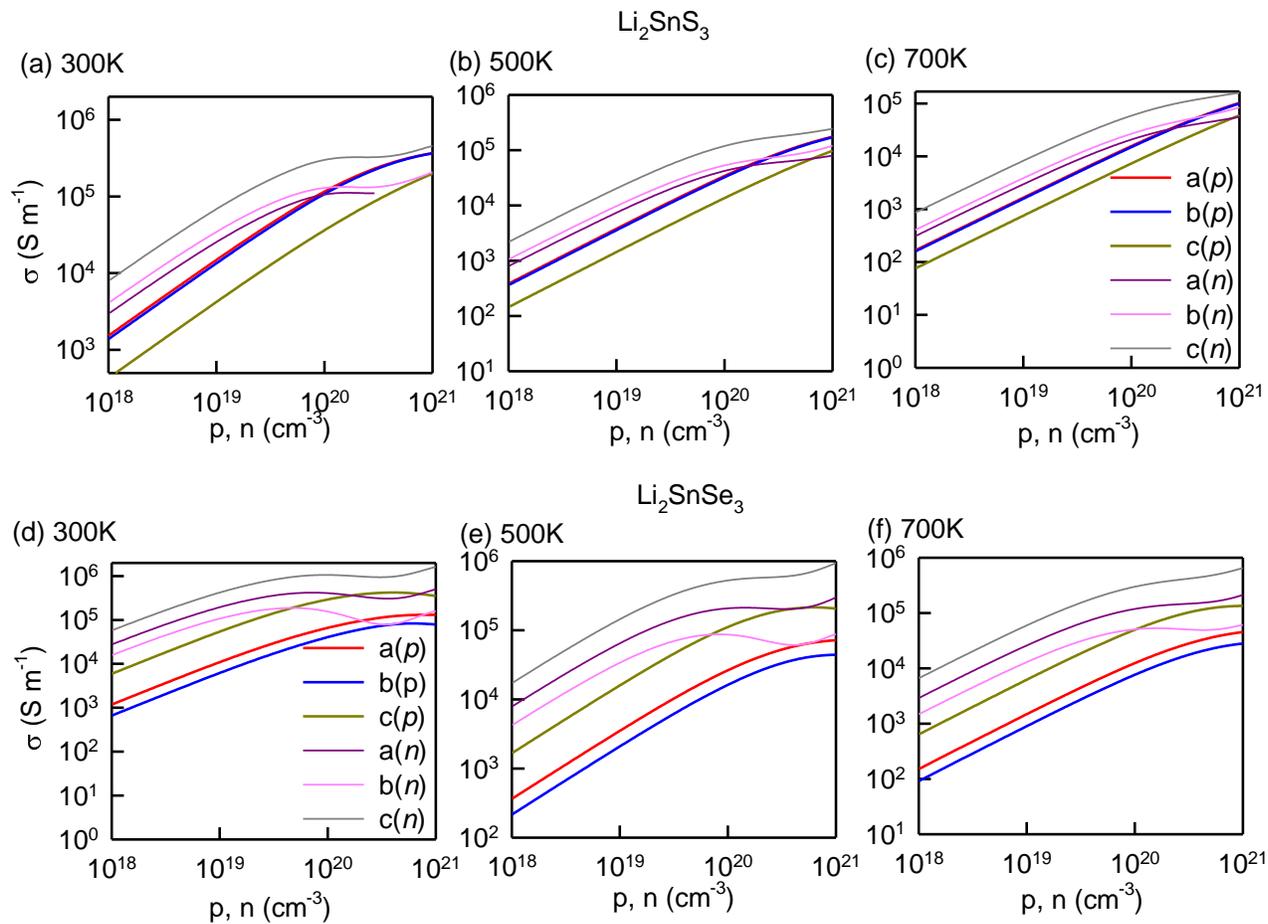

Fig. S1. Anisotropic electrical conductivity of Li$_2$SnS$_3$ (top panel) and Li$_2$SnSe$_3$ (bottom panel) as a function of carrier concentration.

The computed electronic part of the thermal conductivity of both compounds is shown in Fig. S2.

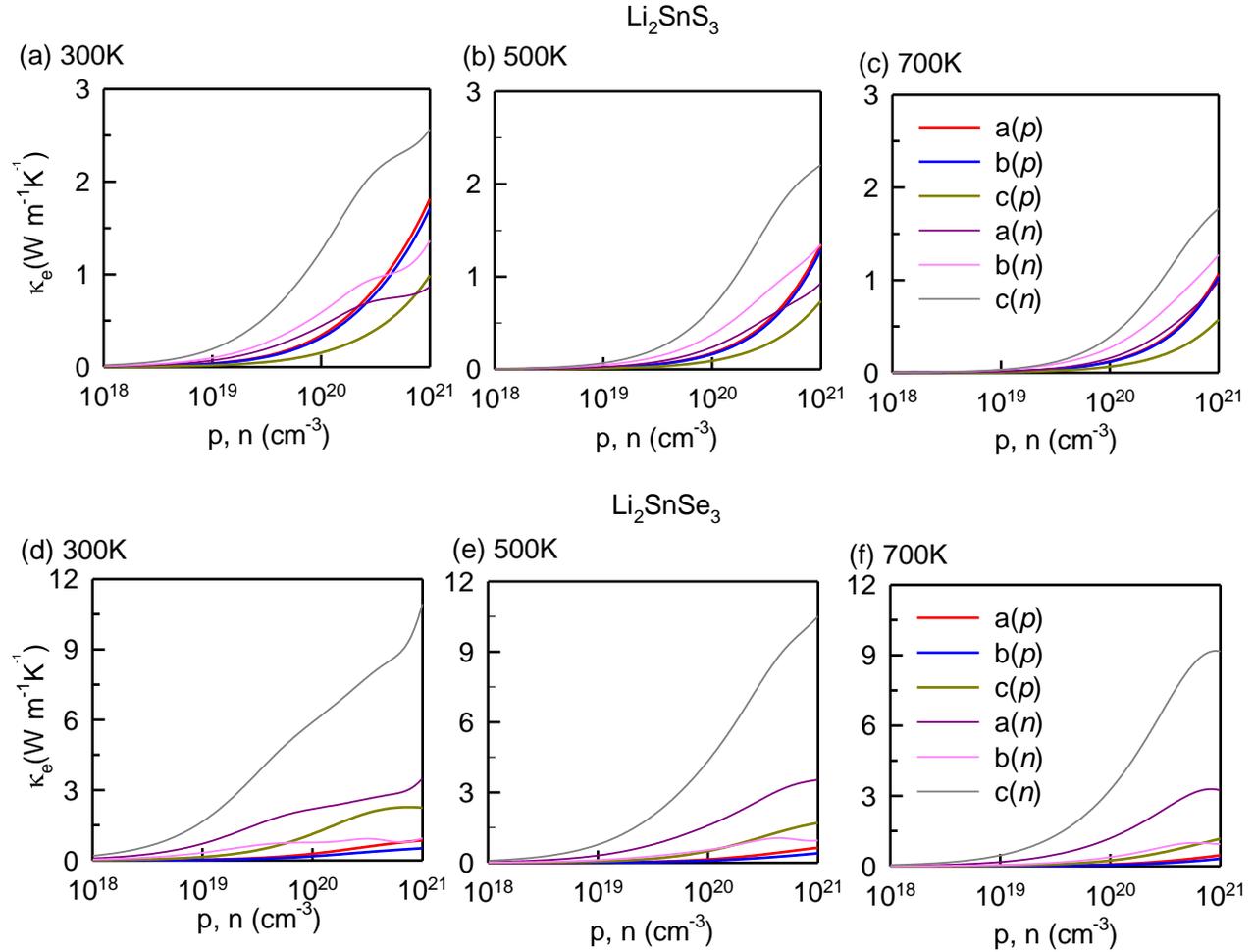

Fig. S2. Carrier concentration dependent of anisotropic electronic part of the thermal conductivity of $Li_2SnS_3$ (top panel) and $Li_2SnSe_3$ (bottom panel). The symbols a, b, and c indicate the crystallographic directions (x, y, z) and $p$, and $n$ indicate the type of carrier.

At high carrier concentration, both compounds exhibit anisotropic electronic thermal conductivity. Although electronic thermal conductivity of $Li_2SnS_3$ is lower than that of $Li_2SnSe_3$, the lattice thermal conductivity is higher, resulting in the suppression of $zT$.